\documentclass[journal]{IEEEtran}
\usepackage{latexsym}
\usepackage{algorithm,algorithmic}
\usepackage{array}
\usepackage{mdwmath}
\makeatletter
\newcommand*{\rom}[1]{\expandafter\@slowromancap\romannumeral #1@}
\usepackage{mdwtab}
\usepackage{eqparbox}
\usepackage{stfloats}
\usepackage{latexsym}
\usepackage{amssymb}
\usepackage{amsmath}
\usepackage{graphicx}
\usepackage{csquotes}
\usepackage{balance}
\usepackage[justification=centering]{caption}
\DeclareGraphicsExtensions{.eps,.pdf,.gif,.png,.jpg}
\usepackage{color}
\usepackage[breaklinks=true]{hyperref}
\usepackage{breakcites}
\usepackage{url}  
\usepackage{amsthm}
\usepackage{multirow}
\usepackage{color}
\usepackage{epstopdf}
\usepackage{endnotes}
\usepackage{framed}
\usepackage{cool} 
\usepackage{enumerate} 
\usepackage{url}
\usepackage[subnum]{cases}
\usepackage[nocomma]{optidef}
\usepackage{etoolbox}
\usepackage{cite} 
\usepackage{caption}
\usepackage{subcaption}

\ifCLASSINFOpdf
\else
\fi
\begin{document}
\title{Trajectory Optimization and Phase-Shift Design in IRS Assisted UAV Network for High Speed Trains}
%
%
%

\author{Yu~Min~Park, Yan~Kyaw~Tun,~
       Zhu~ Han,~\IEEEmembership{Fellow,~IEEE,~}
	   and~Choong~Seon~Hong,~\IEEEmembership{Senior~Member,~IEEE}
	   
\thanks{Yu Min Park, Yan Kyaw Tun, and Choong Seon Hong  are with the Department of Computer Science and Engineering, Kyung Hee University,  Yongin-si, Gyeonggi-do 17104, Rep. of Korea, e-mail:{\{ yumin0906, ykyawtun7, cshong\}@khu.ac.kr}.}
\thanks{Zhu Han is with the Electrical and Computer Engineering Department,
    	University of Houston, Houston, TX 77004, and the Department of
    	Computer Science and Engineering, Kyung Hee University, Yongin-si,
    	Gyeonggi-do 17104,  Rep. of Korea, email{\{zhan2\}@uh.edu}.}}

\maketitle
\vspace{-0.7cm}
\begin{abstract}
The recent trend towards the high-speed transportation system has spurred the development of high-speed trains (HSTs). However, enabling HST users with seamless wireless connectivity using the roadside units (RSUs) is extremely challenging, mostly due to the lack of line of sight link. To address this issue, we propose a novel framework that uses intelligent reflecting surfaces (IRS)-enabled unmanned aerial vehicles (UAVs) to provide line of sight communication to HST users. First, we formulate the optimization problem where the objective is to maximize the minimum achievable rate of HSTs by jointly optimizing the trajectory of UAV and the phase-shift of IRS. Due to the non-convex nature of the formulated problem, it is decomposed into two subproblems: IRS phase-shift problem and UAV trajectory optimization problem. Next, a Binary Integer Linear Programming (BILP) and a Soft Actor-Critic (SAC) are constructed in order to solve our decomposed problems. Finally, comprehensive numerical results are provided in order to show the effectiveness of our proposed framework.
\end{abstract}


%
\IEEEpeerreviewmaketitle

\vspace{-0.5cm}
\section{Introduction}
\vspace{-0.2cm}
\IEEEPARstart{H}{igh} speed trains (HSTs) have gained significant interest from both academia and industry due to their appealing nature towards the realization of fast transportation systems by providing speed up to $200$ km/h. The roadside units(RSUs) can be used to enable HST users' communication for various smart services (e.g., augmented reality). However, RSUs-based communication might not be able to provide true seamless connectivity to HST users due to the frequent handovers because of high speed \cite{b1}. Therefore, there is a need to propose efficient communication architecture for HSTs.

On the other hand, many recent works have deployed unmanned aerial vehicles (UAVs) to enable on-demand, lost cost, and flexible communication infrastructure for end-users. Although UAV-enabled communication infrastructure can offer several benefits, it has a few striking challenges. These challenges are trajectory optimization, limited energy, and resource allocation. The works in \cite{b2} and \cite{b3} primarily focused on UAVs-enabled mobile edge computing and UAVs-assisted HSTs, respectively. Although UAVs can provide line of sight communication to HST users. However, enabling continuous coverage to HST users will require a large number of UAVs, and thus increases the overall cost of the system. Therefore, there is a need to design a cost-effective, efficient UAVs-based communication for HSTs.

Moreover, wireless communication technology through Intelligent Reflecting Surface (IRS) is being studied, recently. IRS is composed of several reflectors, and the angle of the corresponding reflector can be corrected through an electrical signal \cite{b4}. The wireless channels between transmitters and receivers can be flexibly reconfigured using IRS in a wireless network to achieve desired realizations and/or distributions. IRS communication faces many challenges, including resource allocation and phase-shift optimization according to the communication target area. The work in \cite{b5} considered the number of suitable Reflectors to provide IRS-based communication to multi-users. In \cite{b6, cao2021reconfigurable}, the authors approached the problem of optimizing both the power of beamforming and the phase-shift of IRS in a MIMO environment in a numerical manner. However, IRS in a fixed location has a limitation in providing communication as the receiver moves.

IRS assisted UAV network has drawn significant research attention recently. In \cite{b7} and \cite{b8}, the authors tried to overcome communication shading regions by applying IRS located in buildings and UAV-BS together. Though, IRSs located in buildings support passive communication to UAV-BS. In \cite{b9}, the authors used an IRS-attached form of communication infrastructure on UAVs called Aerial Intelligent Reflecting Surface. However the work in \cite{b9} used decomposition and numerical methods to optimize UAVs trajectory and IRS phase-shift that might not be able to respond instantly to a rapidly changing environment of HST. Therefore, we apply reinforcement learning to quickly optimize the trajectory of UAV. Also, we perform the phase-shift optimization with a Transportation Problem (TP) between reflector and HSTs to maximize the minimum data rate of HSTs.

\textcolor{black}{As a result, we propose a method to utilize UAVs and IRS for HST wireless communication. Due to the usage of IRS, UAVs that are slower than HST can extend the communication range even farther. In addition, UAVs' flexibility allows for obstacle avoidance and effective resource utilization.} The main contribution of this paper is to jointly optimize the trajectory of UAV and phase-shift of IRS for maximizing the minimum achievable rate of HSTs. Thus, we introduce that network models through beamforming and IRS required for problem formulation. The formulated problem is non-covex and NP-hard due to the variables of UAV trajectory and IRS phase-shift being coupling. To solve the joint optimization problem, we decompose it into two subproblems. Next, we employ a Binary Integer Linear Programming (BILP) model and reinforcement learning, in order to solve decomposed subproblems. In the simulation results, we validated our proposed method in two aspects. First, the outperformance of the proposed method could be confirmed by comparing the minimum data rate of HSTs when using the fixed IRS and random IRS. Second, we demonstrate that trajectory optimization outperforms a UAV with a fixed height.

\section{System Model and Problem Formulation}

\begin{figure}[h]
\includegraphics[width=7.8cm]{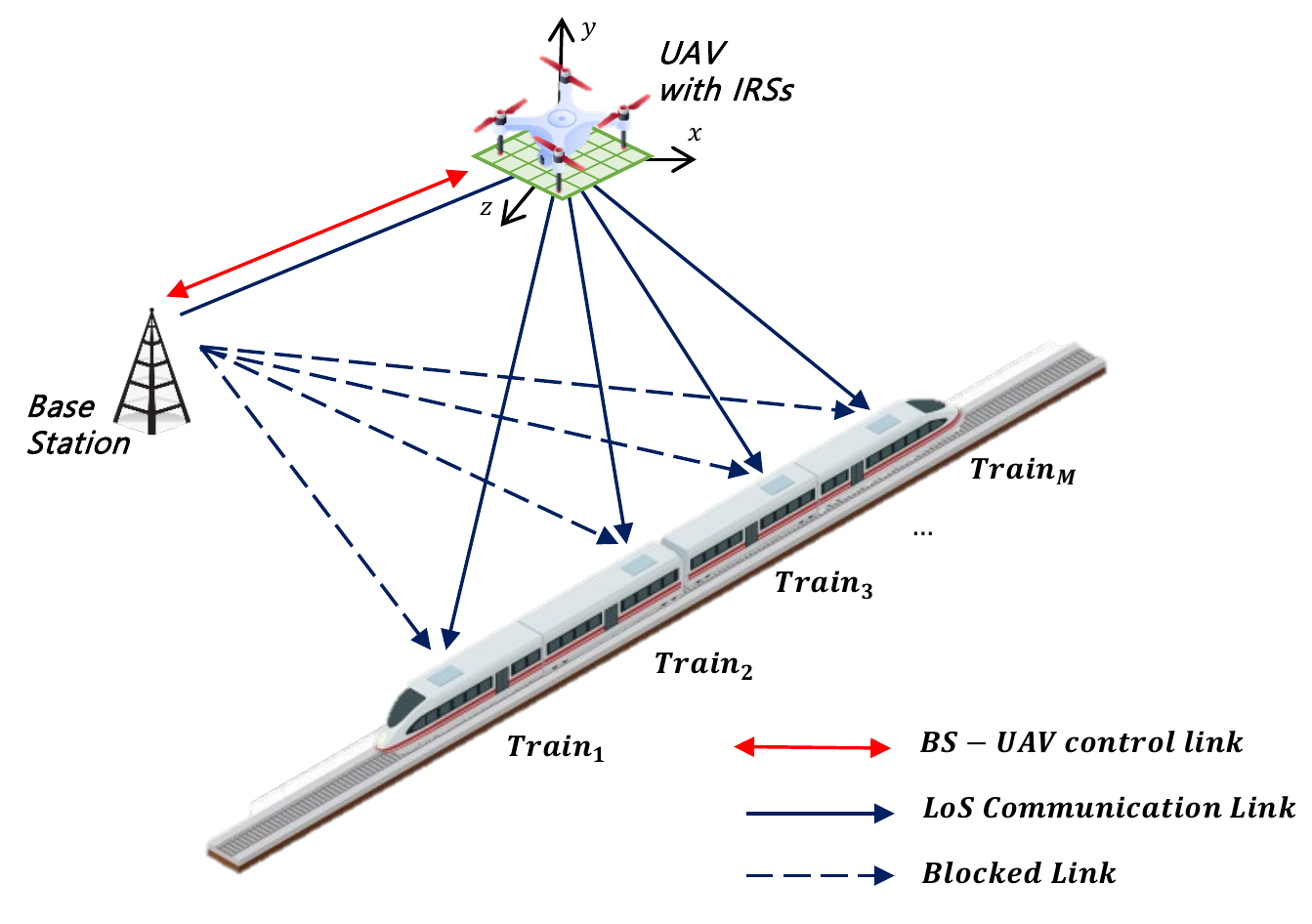}
\caption{System model.}
\label{fig5}
\end{figure}

In Fig.~\ref{fig5} over time $\mathcal{K}=\left\{1,2,...,K\right\}$, we consider a single-input multiple-output (SIMO) wireless system in which a BS equipped with a single antenna serves a set $\mathcal{M}$ of $M$ HSTs through a UAV with IRS. We assume that wireless communication between BS and HSTs is blocked by obstacles, such as buildings. The UAV is equipped with the smart controller that can perform phase-shift of IRS. \textcolor{black}{In addition, we assume that the channel state information (CSI) of all channels involved is perfectly known at the BS and the UAV. Moreover, the free-space path loss  model is adopted because a rural environment was considered.} All HSTs move at the same speed $v_{\mathrm{train}}$ and the speed $v_{u}$ of UAV is limited to $v_{\mathrm{max}}$. Also, UAV $u$ is limited to the height $l_{u}$ between $l_{\mathrm{min}}$ and $l_{\mathrm{max}}$. The IRS mounted on UAV is composed of a set $\mathcal{N}$ of $N$ reflectors, and the normalized power radiation pattern in each reflector can be expressed as \cite{b10}.

\begin{equation}
F(\theta,\varphi)=
\begin{cases}
\cos^3\theta, & \theta \in [0,\frac{\pi}{2}] , \varphi \in [0,2\pi], \\
0, & \theta \in (\frac{\pi}{2},\pi] , \varphi \in [0,2\pi],
\end{cases}\label{eq1}
\end{equation}
where $\theta$ and $\varphi$ are the elevation and azimuth from the antenna to a certain transmitting/receiving direction, respectively.

Let $\Theta\overset{\vartriangle}{=}\left\{\theta_{1}, \theta_{2},..., \theta_{N}\right\}$ denote the ideal IRS phase-shift vector. We define the practical diagonal IRS phase-shift matrix $\Phi$ by multiplying the indicator matrix $\textbf{I}$ for the association between a IRS reflector and HSTs. The indicator matrix $\textbf{I}$ is defined as

\begin{equation}
\textbf{I} = \begin{bmatrix}
i_{1,1} & \cdots & i_{1,M} \\
\vdots & \ddots & \vdots \\
i_{N,1} & \cdots & i_{N,M}
\end{bmatrix},
\label{eq2}\end{equation}
where $i_{n,m}$ is a binary decision variable, i.e., $i_{n,m}\in{\left\{0,1\right\}}, \forall{n\in{\mathcal{N}}}, \forall{m\in{\mathcal{M}}}$. If HST $m$ is associated with a IRS indicator $n$, $i_{n,m} = 1;$ otherwise, $i_{n,m} = 0$. The practical diagonal IRS phase-shift matrix $\Phi$ is defined as

\begin{equation}
\Phi(\textbf{I})=\alpha \mbox{diag}(e^{j\theta_{1}\otimes{\tilde{i}_{1}}}, e^{j\theta_{2}\otimes{\tilde{i}_{2}}}, ..., e^{j\theta_{N}\otimes{\tilde{i}_{N}}}),
\label{eq3}\end{equation}
where $\alpha$ is the fixed amplitude reflection coefficient and $\tilde{i}_{n}=[i_{n,1}, i_{n,2}, ..., i_{n,M}]$ represents the IRS indicator vector of reflector $n$. The achievable channel gain based the distance-dependent path loss over the time slot $k$ is modeled as follow:
\begin{equation}
h(k)=\rho_{0}\left({d_{k}\over{d_{0}}}\right)^{-\delta},
\label{eq7}\end{equation}
where $d_{k}, \rho_0$ and $\delta$ denote the distance between elements over time slot $k$, the path loss at the reference distance $d_0$ and the path loss exponent, respectively \cite{b11}.The deterministic channel from the BS $b$ to the UAV $u$ is denoted by $h_{bu}$. The channel between the UAV $u$ and the HST $m\in M$ is denoted by $h_{um}$. Therefore, The channel from the BS to the HST is defined by $h_{bu}\Phi h_{um}$. The received signal to noise ratio (SNR) at HST $m$ at time slot $k$ is given by
\begin{equation}
\gamma_{m}(k)={{p|F(\theta_t, \varphi_t)F(\theta_r, \varphi_r)h_{bu}(k)\Phi h_{um}(k)|^{2}}\over{\sigma^{2}}},
\label{eq5}\end{equation}
where $\sigma^2$ is the noise power. \textcolor{black}{$\theta_{t}, \varphi_{t}, \theta_{r},$ and $\varphi_{r}$ represent the elevation angle and the azimuth angle from the BS to  the center of the IRS, the elevation angle and the azimuth angle from the center of the IRS to HST $m$, respectively \cite{b10}.} We can express the achievable rate at HST $m$ at time slot $k$ can be written as
\begin{equation}
R_{m}(k)=B\log_{2}(1+\gamma_{m}), \ \ \ \forall m \in {\mathcal{M}},\forall k \in {\mathcal{K}},
\label{eq6}\end{equation}
where $B$ is the BS bandwidth.

Given the network model, our objective is to solve the problem of joint trajectory $\mathcal{T} = \left\{T_{1}, T_{2}, ..., T_{K}\right\}$ of UAV and phase-shift $\mathcal{P} = \left\{\Phi_{1},\Phi_{2}, ..., \Phi_{K}\right\}$ of IRS for wireless HSTs communication with the goal of maximizing the minimum data rate of HSTs over $K$ time slots. We can formulate this problem as follows:

\vspace{-0.2cm}
   \begin{maxi!}[2]
{\mathcal{T}, \mathcal{P}}
{\sum_{k=1}^K{\bar{R}(k)}\label{eq9}}
{\label{eq9}}{}
\addConstraint{\mathrm{C1}: \bar{R}(k)}{=\min_{m\in{\mathcal{M}}}R_{m}(k), \ \ \ \forall k \in {\mathcal{K}} \label{eq9b}}
\addConstraint{\mathrm{C2}: R_{m}(k)\geq{R_0}, \ \ \ \forall{m\in{\mathcal{M}}}, \forall k \in {\mathcal{K}}\label{eq9c}}
\addConstraint{\mathrm{C3}: \Phi_{k}(\textbf{I}_k)=\alpha \mbox{diag}(e^{j\theta_{1}\otimes{\tilde{i}_{1}}}, e^{j\theta_{2}\otimes{\tilde{i}_{2}}}, ..., e^{j\theta_{N}\otimes{\tilde{i}_{N}}}) \label{eq9d}}
\addConstraint{\mathrm{C4}: \tilde{i}_{n}=[i_{n,1}, i_{n,2}, ..., i_{n,M}], \ \ \ \forall n \in \mathcal{N}}  \label{eq9e} 
\addConstraint{\mathrm{C5}: 0 < \theta_n < \frac{\pi}{2}, \ \ \ \forall n \in \mathcal{N}}  \label{eq9f} 
\addConstraint{\mathrm{C6}: 0\leq{v_{u}(k)}\leq{v_{\mathrm{max}}} \ \ \ \forall k \in {\mathcal{K}}}{\label{eq9g}}
\addConstraint{\mathrm{C7}: l_{\mathrm{min}}\leq{l_{u}(k)}\leq{l_{\mathrm{max}}} \ \ \ \forall k \in {\mathcal{K}}}{\label{eq9h}}
\addConstraint{\mathrm{C8}: i_{n,m} \in \left\{0, 1\right\} \ \ \ \forall n \in {\mathcal{N}}, \forall m \in {\mathcal{M}}}{\label{eq9i}},
 \end{maxi!}
where $R_0$ is the minimum achievable rate required by a HST. Here, $\mathrm{C2}$ guarantees a minimum achievable rate for all HSTs. $\mathrm{C3}$ and $\mathrm{C4}$ represents that a phase-shift $\Phi$ is determined through a indicator matrix $\textbf{I}$. $\mathrm{C5}$ ensures that the angles of reflection are between 0 and $\pi/2$. $\mathrm{C6}$ and $\mathrm{C7}$ are for the UAV's behavioral constraints, respectively, for speed and altitude. Additionally, since this paper is to derive the optimal trajectory of the UAV, we assumed that there are sufficient mobile batteries for the UAV.
\vspace{-0.3cm} 
\section{Proposed solution}
From (7), the optimization variables of UAV trajectory and IRS phase-shift are coupling in both objective function and the constraints. It can be shown that the problem becomes non-convex and NP-hard. Therefore, it is challenging to solve our formulated problem in polynomial time. For this reason, we decompose the problem into two subproblems: the optimal phase-shift design for IRS and the UAV trajectory optimization problem in the following two subsections, respectively.
\vspace{-0.3cm}
\subsection{Optimal Phase-Shift of IRS at the given UAV Trajectory}
We discuss about the optimal phase-shift design of IRS at the given UAV trajectory. With the objective of maximizing the minimum data rates of HSTs over the time slot $k$, the optimal phase-shift design of IRS is formulated as follows:
\vspace{-0.3cm} 
\begin{maxi!}|s|[2]<b>
{{\mathcal{P}_{k}}}{\bar{R}(k)\label{eq10}}
{\label{eq10}}{}
\addConstraint{(\mathrm{C1})\text{-}(\mathrm{C5}),(\mathrm{C8}) \label{eq10b}}.
\end{maxi!}

Since the interference between reflectors is not considered in this paper, we can formulate problem (8) of finding the optimal indicator matrix as TP. Here, each reflector is a source and each HST is a destination. \textcolor{black} {Therefore, the proposed subproblem can be solved by using a BILP with indicator matrix $\textbf{I}$ composed of decision variables $i_{n,m}$. To apply the BILP method, subproblem (8) can be reformulated by maximizing an additional variable $Z$ that is an upper bound for the achievable rates $R_{m}$ of each HST $m$ as follows:}
\vspace{-0.3cm}
\begin{maxi!}|s|[2]<b>
{{\textbf{I}}}{Z\label{eq11}}
{\label{eq11}}{}
\addConstraint{\sum_{m\in{\mathcal{M}}}i_{n,m}=1, \ \ \ \forall{n\in{\mathcal{N}}}\label{eq11b}}
\addConstraint{R_{m}=\sum_{n\in{\mathcal{N}}}{i_{n,m}\cdot R_{n,m}}, \ \ \ \forall{m\in{\mathcal{M}}}\label{eq11c}}
\addConstraint{Z\leq R_{m}, \ \ \ \forall{m\in{\mathcal{M}}}, \label{eq11d}}
\end{maxi!}
where \eqref{eq11b} ensures that  a single IRS reflector is assigned to only one HST. \textcolor{black} {In \eqref{eq11c}, $R_{n,m}$ is the achievable rate between IRS reflector $n$ and HST $m$. Therefore,} \eqref{eq11c} represents that HST's achievable rate is the sum of achievable rates from associated IRS reflectors. \textcolor{black} {we can solve the subproblem (9) using GEKKO as an optimization tool for BILP \cite{b15}.}
\vspace{-0.3cm}
\subsection{Optimal Trajectory of UAV at the given IRS Phase-shift}

Based on the optimal phase-shift of the IRS over time slot $k$, the optimal trajectory of the UAV is derived. Therefore, the optimization trajectory of UAV at the given phase-shift of IRS can be expressed as follows:
\vspace{-0.3cm}
\begin{maxi!}|s|[2]<b>
{{\mathcal{T}_{k}}}{\bar{R}(k)\label{eq12}}
{\label{eq12}}{}
\addConstraint{(\mathrm{C1}),(\mathrm{C2}),(\mathrm{C6}),(\mathrm{C7}) \label{eq12b}}.
\end{maxi!}

To solve problem (10), we deploy the Soft Actor-Critic (SAC) method \cite{b13}. \textcolor{black}{SAC is an off-policy model that trains into the Replay Buffer at the same time as a variation of the Actor Critic model, which is a policy gradient method. SAC is efficient for optimizing the motion of drones or robots in continuous spaces, such as the proposed problem in (10).} The advantage of using reinforcement learning over numerical method is that the optimal value can be derived by entering the status in real time. The episodes of reinforcement learning are divided into certain times $\mathcal{K}^\prime=\left\{1^\prime,2^\prime,...,K^\prime\right\}$. The elements of reinforcement learning are designed to maximize the minimum achievable rate of HSTs. The three basic elements of reinforcement learning are as follows: state, action, reward.

In time slot $k^\prime$, state $s_{k^\prime}$ is defined as
\vspace{-0.3cm}
\begin{equation}
s_{k^\prime} = \left\{p_{b,k^\prime},p_{u,k^\prime},p_{\mathcal{M},k^\prime}\right\},
\label{eq13}\end{equation}
where $p_{b,k^\prime}, p_{u,k^\prime}$ and $p_{M,k^\prime}$ are the positions of the BS $b$, UAV, and a set $\mathcal{M}$ of HSTs in time slot $k^\prime$. That's when the heights of the base sation and HSTs are always 0. In other words, the BS and HSTs are located only on the ground. 

The action is the amount of movement in the x-axis, y-axis and z-axis to move from the current position in time slot $k^\prime$,
\vspace{-0.2cm}
\begin{equation}
a_{k^\prime} = \left\{\bigtriangleup x_{u,k^\prime}, \bigtriangleup l_{u,k^\prime},\bigtriangleup z_{u,k^\prime}\right\}.
\label{eq14}\end{equation}

The reward over time $k^\prime$ is divided into three and can be expressed as follows:
\vspace{-0.3cm}
\begin{equation}
r_{k^\prime}=
\begin{cases}
0, & \mbox{if }\bigtriangleup{\bar{R}}\leq{0}\mbox{ or }\bar{R}<R_{0}, \\
\bigtriangleup{\bar{R}}, & \mbox{if }\bigtriangleup{\bar{R}}>0, \\
-1, & \mbox{if } l_{u,k^\prime} > l_{\mathrm{max}} \ \mbox{or} \ l_{u,k^\prime} < l_{\mathrm{min}},
\end{cases}
\label{eq15}\end{equation}
where $\bigtriangleup{\bar{R}}$ denotes the change in $\bar{R}$, i.e., $\bigtriangleup{\bar{R}}=\bar{R}-\bar{R}_{\mathrm{pre}}$. First, if $\bar{R}$ becomes larger after the action than the current $\bar{R}$, it is rewarded as much as the increased amount and the episode continues. Second, the episode ends when $\bar{R}$ is less than the minimum required amount $R_0$ or less than the previous $\bar{R}_{\mathrm{pre}}$. Lastly, when the UAV's altitude is out of the limited altitude, the episode ends with a disreward, $-1$.

Algorithm 1 summarizes our proposed optimal strategy for IRS assisted UAV network. Inputs are HSTs position $P_{\mathcal{M}}$ over the time interval $\mathcal{K}$ and BS position $p_{b}$, and the UAV's initial position $p_{u,0}$. First, we derive the optimal indicator matrix $I^{*}_k$ for the current position by the BILP method and perform the optimal phase-shift $\Phi^{*}_k$ based on it. Based on the pre-learned model, the optimal UAV Trajectory $T^{*}_k$ is derived. Repeat every time interval $k$ like this. Consequently, we can derive the optimal trajectory $\mathcal{T}$ and phase-shift $\mathcal{P}$ for all time $\mathcal{K}$.

\begin{algorithm}[t]
	\caption{Optimal Strategy for UAV with IRS}
	\begin{algorithmic}[1]
		\renewcommand{\algorithmicrequire}{\textbf{Input:}}
		\renewcommand{\algorithmicensure}{\textbf{Output:}}
		\REQUIRE $p_{u,0}$, $p_{b}$,$P_{\mathcal{M}}= \left\{p_{\mathcal{M},1}, p_{\mathcal{M},2}, ...,p_{\mathcal{M},K} \right\}$, $T_{max}$
		\FOR {$k$ in $\left\{1,2,...,K   \right\}$}
		\FOR {$t$ in $T_{max}$}
		\STATE Phase-shift $\Phi^{*}_k$ of IRS based on the optimal indicator matrix $I^{*}_k$ by BILP in (9)
		\STATE Trajectory $T^{*}_k$ of UAV based on reinforcement learning model in (10)
		\ENDFOR
		\ENDFOR
		\ENSURE $\mathcal{T}, \mathcal{P}$
	\end{algorithmic}
\end{algorithm}

\vspace{-0.3cm}
\section{Simulation results}

\begin{table}
	\caption{\textbf{Parameters for Simulation.}}
	\label{table}
	\setlength{\tabcolsep}{2.8pt}
	{\footnotesize
		\renewcommand{\arraystretch}{1.4}
		\begin{tabular}{|p{60pt}| p{120pt}| p{55pt}|}
			\hline
			\textbf{Parameters} & \textbf{Description}  & \textbf{Value}	\\
			\hline
			$\rho_{0}, d_0$&the path loss at the reference distance&-20 dB, 1 m\\
			\hline
			$B$&The bandwidth at the base station&20 MHz\\
			\hline
			$\sigma^{2}$&The noise power&-100 dBm\\
			\hline
			$p$&The transmitted power&10 dBm\\
			\hline
			$\delta_{bu}, \delta_{ut}$&The path loss exponents of base station, UAV, and trains&2.6, 2.8\\
			\hline
			$M$&The number of HSTs&4\\
			\hline
			$N$ &The number of reflectors in IRS & 100 (10 $\times$ 10)\\
			\hline
			$N_{x}, N_{y}$ & $x, y$ interval between reflectors & 0.01 m\\
			\hline
			$v_{train}$&The speed of HSTs&100 m/s\\
			\hline
			$v_{uav}^{max}$&The maximum speed of UAV&55 m/s\\
			\hline
		\end{tabular}
	}
	\label{tab2}
	\vspace{-0.2em}
\end{table}

The ML-agent library is used to use reinforcement learning \cite{b14}. The specifications of the computer used in the simulation are i7-9700K 3.60GHz, 32GB and RTX 2070 SUPER 8GB. During reinforcement learning, random BS and HSTs are assigned at the beginning of every episode. \textcolor{black} {In this paper, we simulate within a communication range that one UAV can cover the speed of HST.} In order to check the simulation result, the HST is moved on a fixed track, and a BS is placed in the center. The speed of the HST used in the experiment moves constantly at $100$ m/s, and it consists of a total of 4 HSTs. The IRS used in the experiment have a total of 100 reflectors, and the interval between reflectors is $0.01$ m. Other definitions and values of variables used for communication can be found in Table ~\ref{tab2}.

Including the proposed joint trajectory and phase-shift, we consider four benchmark algorithms as follows:
\begin{itemize}
    \item \textit{SU with OI}: Algorithm using the optimal phase-shift of IRS with the static UAV position
    \item \textit{OU with RI}: Algorithm using the optimal trajectory of UAV with the random phase-shift of IRS
    \item \textit{OU with SI}: Algorithm using the optimal trajectory of UAV with the static phase-shift of IRS
    \item \textit{OU with OI (proposed method)}: Algorithm using the optimal trajectory of UAV with the optimal phase-shfit of IRS
\end{itemize}

All of the algorithms were run in the same communication environment. In the SU with OI algorithm, the UAV was positioned at the center of the base station and the trajectory of HST, and the height was fixed at $100$ m. And in the OU with SI algorithm, reflectors of IRS were all assigned the same number regardless of the location of HSTs.

\begin{figure}[t]
\centering\includegraphics[width=8.0cm]{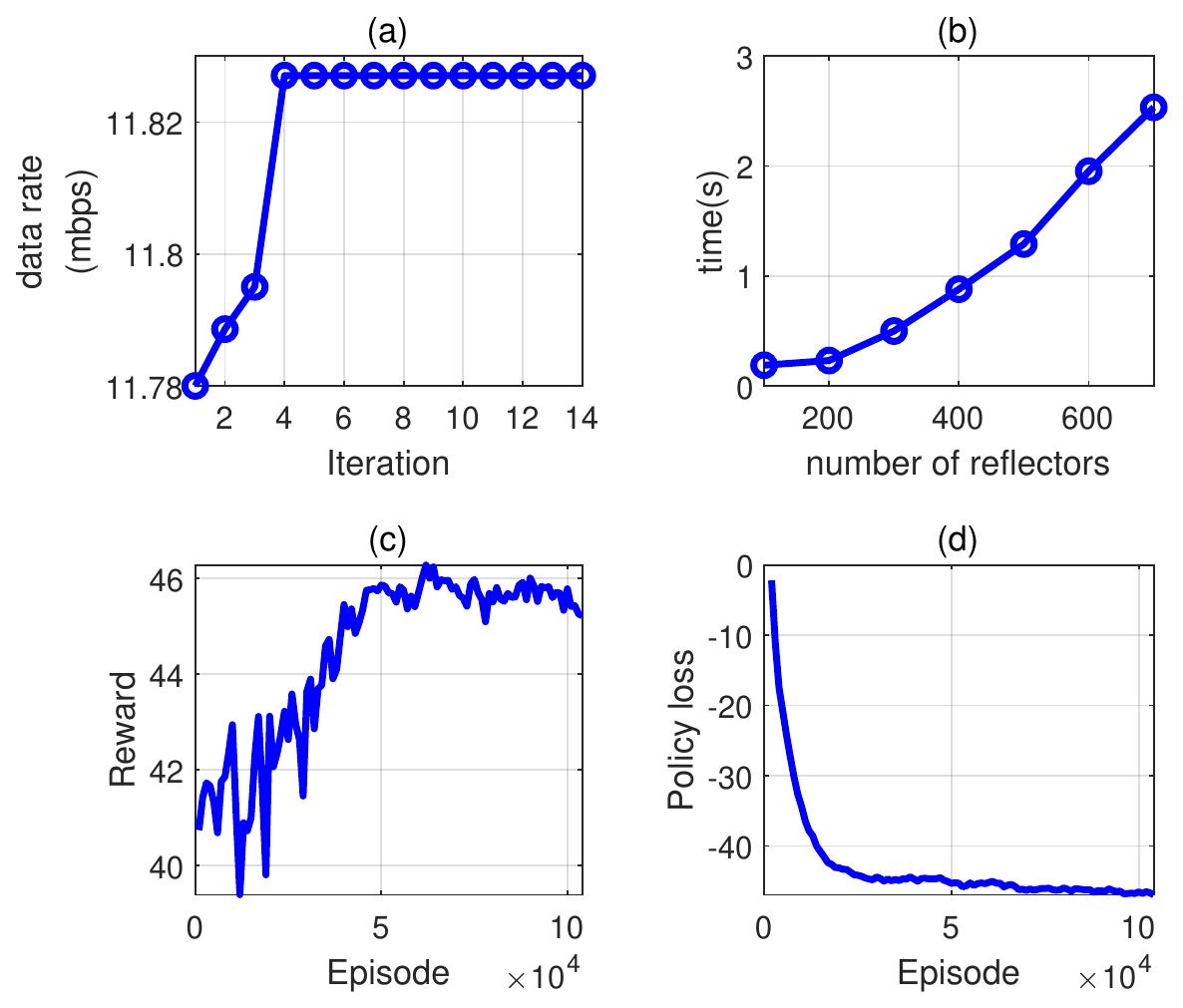}
\caption{The efficiency and results of the algorithm.}
\label{fig7}
\end{figure}

Fig.~\ref{fig7} shows the efficiency and results of the algorithm used for optimization. \textcolor{black}{Fig.~\ref{fig7}(a) shows that the results converge with the iterations of the proposed algorithm.} Fig.~\ref{fig7}(b) is results of the number of reflectors used in the TP for phase-shift. As the number of reflectors increases, it can be seen that the resulting time increases consistently. This shows that the method BILP we used is global optimization and this has a problem in which the computational volume increases consistently as the number of reflectors increases. Thus, we conducted within the number of reflectors that can derive the optimal solution in real-time. Fig.~\ref{fig7}(c) and Fig.~\ref{fig7}(d) are results during reinforcement learning. As we repeat the episode, we can see that the reward increases and the loss decreases. Through this, we can see that the algorithm was able to find the optimal location for increasing the minimum achievable rate.

\begin{figure}[t]
\centering\includegraphics[width=7.8cm]{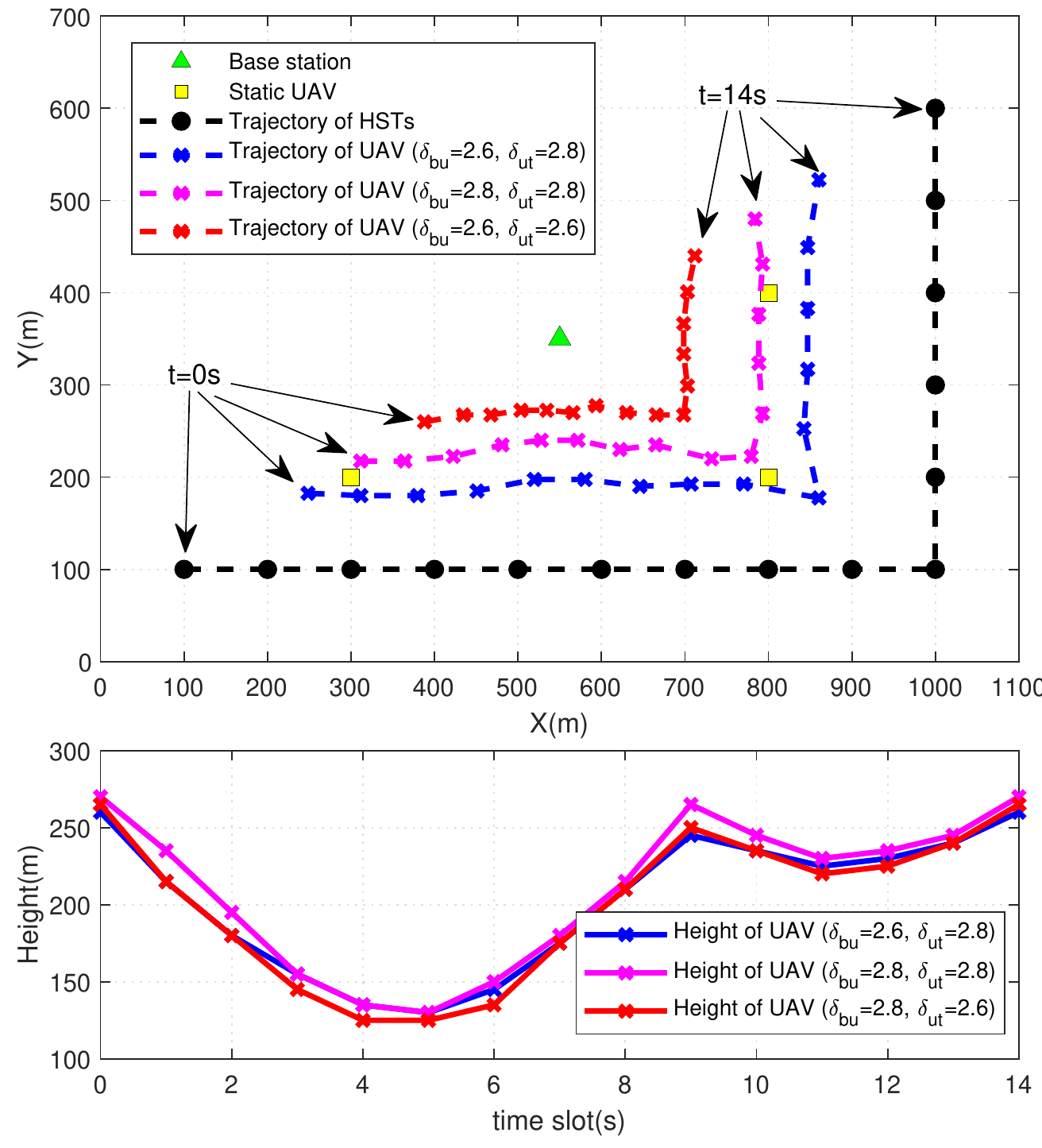}
\caption{Illustration of the test scenario and result.}
\label{fig8}
\end{figure}

Fig.~\ref{fig8} shows a simple scenario for implementing the proposed algorithm measurement. First, the black dotted line denotes a predetermined HSTs path. The red triangle ($550$, $0$, $350$)  plays the role of a BS in the middle. The green squares $[(300, 100, 200), (800, 100, 200), (800, 100, 400)]$ represent the position of the fixed IRS used in the \textit{SU with OI} algorithm. Each route derives the optimal UAV location for the location of HST at 1 second intervals. \textcolor{black}{As a result, the path of the optimal UAV obtained through our proposed algorithm is a dotted line with an blue, magenta, red cross according to path loss exponents.} As for the altitude of the UAV, it is effective to support communication at a low altitude when the HSTs and a BS are close, and at a high altitude when far away. This result is due to the loss that occurs according to the angle when the phase-shift of IRS. \textcolor{black}{Moreover, it can be seen that the difference in altitude according to path loss exponents is not large.} 
\begin{figure}[t]
\centering\includegraphics[width=6.2cm]{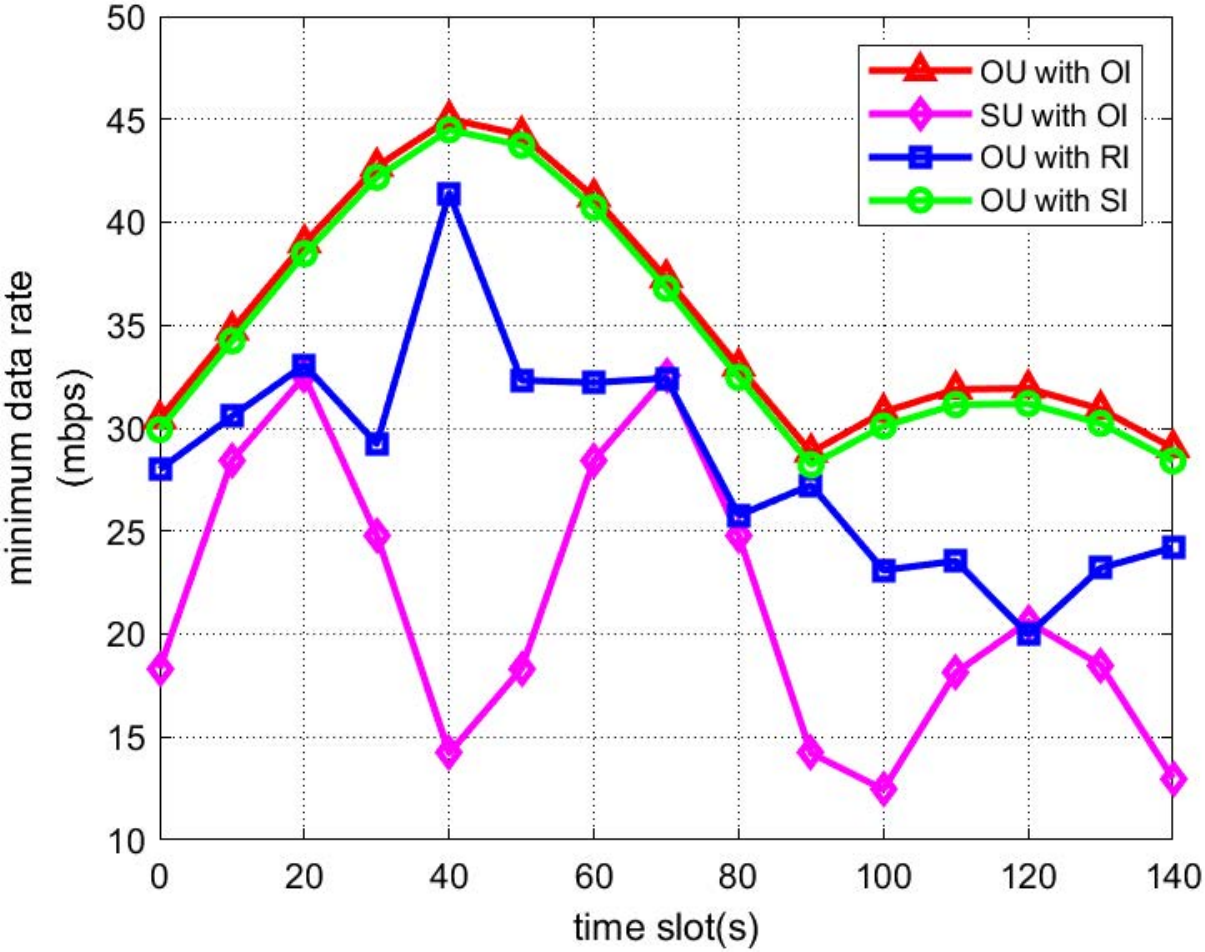}
\caption{The minimum achievable rate by the methods.}
\label{fig9}
\end{figure}

\begin{figure}[t]
\centering
\includegraphics[width=6.2cm]{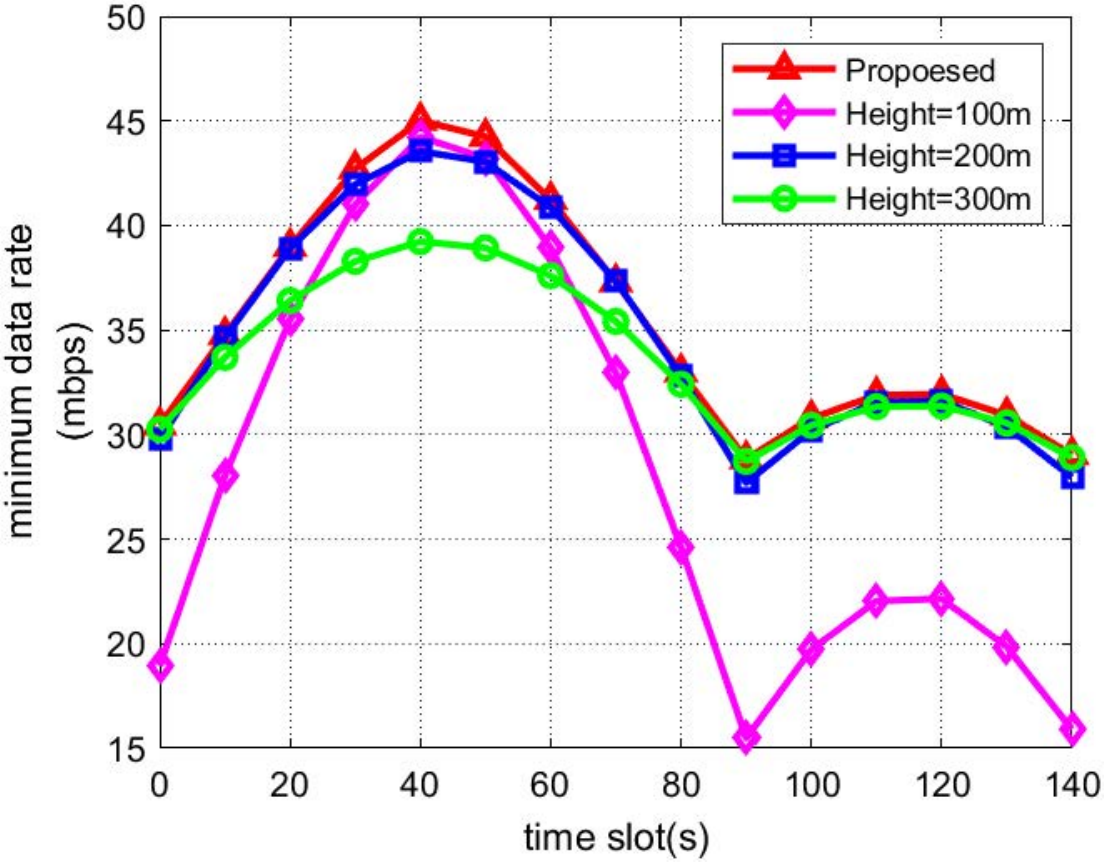}
\caption{The minimum achievable rate by the heights.}
\label{fig10}
\end{figure}
Fig.~\ref{fig9} is a graph showing the minimum achievable rate for each method over time. The proposed algorithm, \textit{OU with OI}, shows the best results. The lowest performance is that IRS is randomly determined by \textit{OU with RI}. Next, the case of using a fixed UAV and an optimized IRS is \textit{SU with OI}. This method seems to be the result of not being able to move fluidly according to the positional movement of HST. Next, \textit{OU with SI} is the case of executing the IRS phase-shift in the same manner without considering the location of HST.  From \textit{OU with SI}, we can see the importance of the optimal phase-shift in the optimal strategy. The graph shows that \textit{OU with SI} has the most similar performance compared to the proposed method \textit{OU with OI}. It can be concluded that UAV trajectory is a more important factor than IRS phase-shift in the optimal strategy. Also, we can predict that the more reflectors of the IRS, the greater the effect of the optimal IRS phase shift. In the optimized result over time slots, it can be seen that the minimum achievable rate changes according to the distance from the HSTs.

Fig.~\ref{fig10} is a graph of the data rate according to the altitude when the IRS phase-shift is optimized. When the altitude is 100m and 300m, the performance is poor, and when it is fixed at 200m, it is close to the proposed algorithm, but shows a low value. As a result, it can be confirmed that the altitude is optimized by considering the locations of the base station and HSTs through the proposed algorithm. 
\vspace{-0.3cm}
\section{Conclusion}
In this paper, we proposed a novel IRS assisted UAV framework to provide stable communication services for HSTs. Then, we have investigated joint UAV trajectory optimization and optimal phase-shit design of IRS of the proposed model with the aim of maximizing the minimum achievable data rates of HSTs. To be tractable, the formulated problem is further decomposed into two subproblems. Later, we have applied BILP and SAC in order to solve the decomposed subproblems. In the simulation, we confirm how efficiently our proposed system can support wireless communication for HSTs than stationary IRS or BS. The proposed algorithm derives the maximum {19.9\%} and {4\%} higher data rates, respectively, compared with the fixed IRS and fixed phase-shift.
\vspace{-0.3cm}

\ifCLASSOPTIONcaptionsoff
  \newpage
\fi

\bibliographystyle{IEEEtran}
\bibliography{ref}

\end{document}